\documentclass[12pt,a4paper]{article}
\usepackage[sc,noBBpl]{mathpazo}
\usepackage[mathscr]{eucal}
\usepackage{amsmath,amsfonts,amssymb,amsthm}
\usepackage{graphicx}
\usepackage{mathtools}
\usepackage{color}
\usepackage{tgpagella}
\usepackage{subfigure}
\usepackage{caption}
\usepackage{float}
\newcommand\cH{{\mathcal H}}

\newcommand\cO{{\mathcal O}}

\newcommand\scH{{\mathscr H}}

\newcommand\mvector{\boldsymbol}

\newcommand\vp{\mvector{p}}
\newcommand\vq{\mvector{q}}
\newcommand\vr{\mvector{r}}

\newcommand\vv{\mvector{v}}

\newcommand\vx{\mvector{x}}

\newcommand\vA{\mvector{A}}

\newcommand\vH{\mvector{H}}
\newcommand\vI{\mvector{I}}
\newcommand\vJ{\mvector{J}}
\newcommand\vK{\mvector{K}}
\newcommand\vL{\mvector{L}}
\newcommand\vM{\mvector{M}}
\newcommand\vN{\mvector{N}}

\newcommand\vP{\mvector{P}}
\newcommand\vQ{\mvector{Q}}

\newcommand\vX{\mvector{X}}

\newcommand\vOmega{\mvector{\Omega}}

\newcommand\field{\mathbb}

\newcommand\R{\field{R}}

\newcommand\const{\operatorname{const}}

\newcommand\diag{\operatorname{diag}}

\newcommand\rmi{\mathrm{i}\mspace{1mu}}

\newcommand\pder[2]{\dfrac{\partial #1 }{\partial #2}}

\usepackage{url}

\theoremstyle{plain}
\newtheorem{theorem}{Theorem}

\newtheoremstyle{note}{\topsep}{\topsep}{\slshape}{}{\scshape}{}{ }{}
\theoremstyle{note}

\numberwithin{equation}{section}
\numberwithin{theorem}{section}
\numberwithin{definition}{section}
\numberwithin{lemma}{section}
\numberwithin{proposition}{section}
\numberwithin{corollary}{section}
\numberwithin{remark}{section}

\mathtoolsset{mathic,centercolon}

\usepackage{subfigure}

\title{Analysis of constrained 2-body problem}
\author{
   Wojciech Szumi\'nski$^1$ and Tomasz Stachowiak$^2$ \\[1em]
  {}$^{1}$Institute of Physics, \\ University of Zielona G\'ora, 
  Licealna 9,  \\
  PL-65-407,  Zielona G\'ora, Poland\\
  {}$^2$ Center for Theoretical Physics PAS,\\  Al. Lotnikow 32/46, 02-668 Warsaw, Poland}
\begin{document}
%
\maketitle

\begin{abstract}
We consider the system of two material points that interact by elastic
forces according to Hooke's law and their motion is restricted to certain curves lying
on the plane. The nonintegrability of this system and idea of the proof are communicated.
Moreover, the analysis of global dynamics by means of Poincar\'e cross sections
is given and local analysis in the neighborhood of an equilibrium is performed by
applying the Birkhoff normal form. Conditions of linear stability are determined and
some particular periodic solutions are identified.
\end{abstract}
\date{\small Key words: pendulum; non-integrability; Morales--Ramis theory; differential  Galois theory; Poincar\'e sections; chaotic Hamiltonian systems; Birkhoff normalization; stability analysis.}
\maketitle

\section{Introduction\label{sec:1}}
Seeking exact solutions of nonlinear dynamical systems is a task to which physicists,
engineers and mathematicians have devoted much of their time over the centuries.
But to date, only a few particular examples of real importance have been found. In a
typical situation, nonlinear equations of motion are nonintegrable and hence we have
little or no information about qualitative and quantitative behavior of their solutions.
However, a very useful tool to overcome these difficulties is the so-called Birkhoff
normalization. The idea of this treatment, which is used in Hamiltonian systems,
goes back to Poincar\'e and it was broadly investigated by Birkhoff in~\cite{Birkhoff::27}. It is mainly
based on the simplification of the Hamiltonian expanded as a Taylor series in the
neighborhood of an equilibrium position by means of successive canonical transformations.
Using the Birkhoff normalization one can: determine stability of equilibrium
solutions; find approximation of analytic solutions of Hamiltonian equations;
and identify families of periodic solutions with given winding numbers.

The aim of this paper is the analysis of dynamics of the system of two material
points that interact by elastic force and can only move on some curves lying
on the plane. At first in order to get a quick insight into the global dynamics of the
considered system, we make a few Poincar\'e cross sections for generic values of parameters.
Moreover, we communicate the nonintegrability result and give the idea of
their proof. Next, we look for equilibria of the system and we make the local analysis
in their neighborhood by means of Birkhoff normal form. In particular we look
for values of parameters for which equilibrium is linearly stable, we plot resonance
curves and look for periodic solutions corresponding to them.
\section{Description  of the system and its dynamics \label{sec:2} }
\begin{figure}[h!]
	\vskip -20pt
\centering{
\resizebox{85mm}{!}
{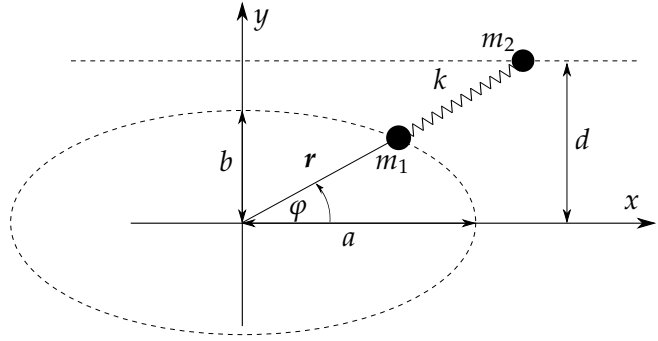}
\caption{Geometry of the system \label{fig:model_1}}
}
\end{figure}
\begin{figure}[H]
	\vskip -10pt
	\centering{
		\resizebox{88mm}{!}
		{\input{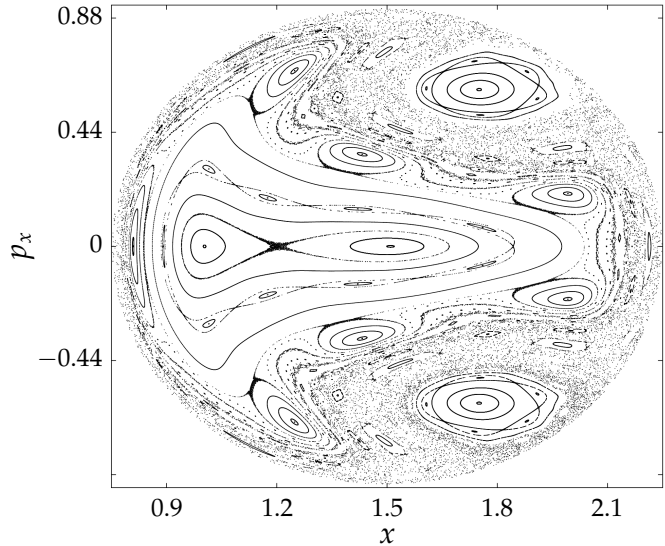}}
		\caption{Poincar\'e cross-section on the surface $\varphi=0$ with $p_\varphi>0$ for the values of parameters: $ E=0.28, \ m_1=1,\ m_2=1.5, \ k=1,\ a=1.5,\ b=1, \ d=0$\label{fig:poin_model_4a}}
	}
\end{figure}
\begin{figure}

	\centering{
		\resizebox{88mm}{!}
		{\input{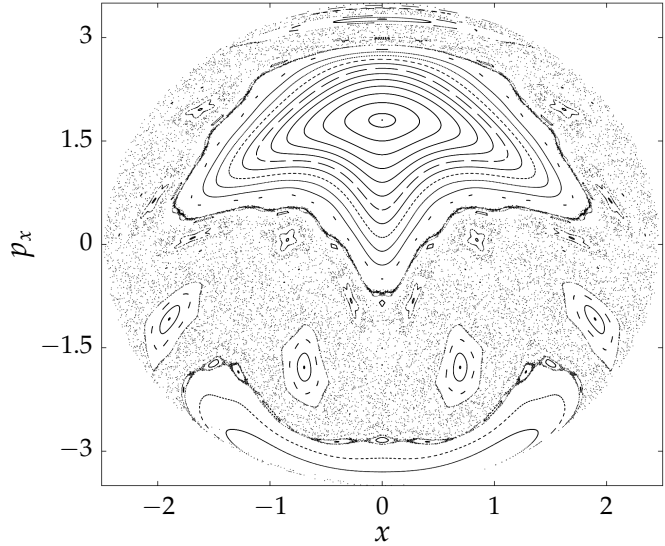}}
		\caption{Poincar\'e cross-section on the surface $\varphi=0$ with $p_\varphi>0$ for the values of parameters: $ E=3.2, \ m_1=1,\ m_2=2, \ k=1,\ a=1,\ b=1.5, \ d=2$\label{fig:poin_model_4b}}
	}
\end{figure}

In Fig.~\ref{fig:model_1} the geometry of the system is shown. It consists
of two masses $m_1$ and $m_2$ connected  by a spring with  elasticity coefficient
$k$. The first mass, $m_1$, moves on an ellipse parametrized by
$\vr=(a\cos\varphi,b\sin\varphi)^T$, while the second one, $m_2$, moves along the
straight line parallel to the $x$-axis and  shifted from it by the distance
$d$. The Lagrange function corresponding to this model is as follows
\begin{equation}
\label{eq:Lagrange1}
L=\frac{1}{2}m_1\left(b^2\cos^2\varphi+a^2\sin^2\varphi \right)\dot\varphi^2+\frac{1}{2}m_2\dot x^2-\frac{1}{2}k\left[(x-a\cos\varphi)^2+\left(d-b\sin\varphi\right){}^2\right].
\end{equation}
In order to have invertible Legendre transformation we assume that the
condition $m_1m_2\neq 0$ is always satisfied. Thus, the Hamiltonian function
can be written as
\begin{equation}
\label{eq:Hamiltonian1}
H=\frac{p_\varphi^2}{2m_1(b^2\cos^2\varphi+a^2\sin^2\varphi)}+\frac{p_x^2}{2m_2}+\frac{1}{2}k\left[(x-a\cos\varphi)^2+(d-b\sin\varphi)^2\right],
\end{equation}
and the equations of motion are given by
\begin{equation}
\begin{aligned}
    \dot x &=\frac{p_x}{m_2}, \quad \dot p_x =-kx+ak\cos\varphi,\quad
    \dot \varphi =\frac{p_\varphi}{m_1(b^2\cos^2\varphi+a^2\sin^2\varphi)},\\ 
 \dot p_\varphi &=\frac{(a^2-b^2)\sin(2\varphi)p_\varphi^2}{2m_1(b^2\cos^2\varphi+a^2\sin^2\varphi)^2}+[bd+(a^2-b^2)\sin\varphi]k\cos\varphi-akx\sin\varphi.
\end{aligned}
\label{eq:vh1}
\end{equation}
In order to present the dynamics of the considered model we made several
Poincar\'e cross sections which are presented in
Figs.~\ref{fig:poin_model_4a}-\ref{fig:poin_model_4b}.
 As we can see,  they show that for generic values of parameters and sufficiently large energies the system
 exhibits chaotic behavior. In fact  we can  prove the following theorem. 
 \begin{theorem}
 \label{th:1}
The system of two point masses $m_1$ and $m_2$, such that $ab(m_1- m_2)\neq0$,
one moving on an ellipse and the other on the straight line containing a semi-axis of the
ellipse, is not integrable in the class of functions meromorphic
in coordinates and momenta.
\end{theorem}
The proof of the above theorem consists in the direct application of the
so-called Morales--Ramis theory that is based on analysis of the differential
Galois group of variational equations. They are obtained by lineralization of
the equations of motion along a particular solution that is not an equilibrium
position. For the precise
formulation of the Morales--Ramis theory and the definition of the
differential Galois group see e.g.,~\cite{Morales::99,Put::03}. The  main
theorem of this theory states that if the system is integrable in the Liouville
sense, then the identity component of the differential Galois group of
variational equations is Abelian, so in particular it is solvable. The  technical details of the proof of this theorem are given in~\cite{Szuminski::16}, where the authors show that
 for this system these necessary
integrability conditions are not satisfied.
\section{Stability analysis - Birkhoff normalisation}
Although Hamiltonian~\eqref{eq:Hamiltonian1} turned out to be not integrable we
can deduce some important information about the dynamics from its Birkhoff normal
form. But first, in order to minimize the number of parameters and thus simplify
our calculations as much as possible, we rescale the variables in $H$~\eqref{eq:Hamiltonian1} in the following way
\begin{equation}
q_1=\frac{\pi}{2}-\varphi,\quad p_1=\frac{\tau }{b^2m_2}p_\varphi,\quad q_2=\frac{x}{b},\quad p_2=\frac{\tau}{bm_2}p_x,\quad e=\frac{\tau^2 }{m_2b^2}E.
\end{equation}
Choosing  $\tau=\omega_0^{-1}$, where $\omega_0=\sqrt{k/m_2}$, the dimensionless Hamiltonian takes the form
\begin{equation}
\label{eq:hamres1}
H=\frac{1}{2}\left(\frac{p_1^2}{\alpha(\beta^2\cos^2q_1+\sin^2q_1)}+p_2^2+\left(x-\beta\sin q_1\right)^2+\left(\delta-\cos q_1\right)^2\right).
\end{equation}
The new dimensionless parameters $(\alpha,\beta,\delta)$ are defined by
\begin{equation}
\label{eq:par1}
\alpha=\frac{m_1}{m_2},\qquad \beta=\frac{a}{b},\qquad \delta=\frac{d}{b}-1.
\end{equation}
Let us denote $\vx=(q_1,q_2,p_1,p_2)^T$, and 
let $\dot{\vx}=\vv_H(\vx)=  \vJ\nabla H$ be the Hamiltonian
vector field generated by the Hamiltonian~\eqref{eq:hamres1}, then it is easy
to verify that the equilibrium  $\dot{\vx}=0$ is localized at the origin. Thus, if we
assume that $H$ is analytic in the neighborhood of  $\vx=0$, then we can
represent it as a Taylor series
\begin{equation}
H=H_2+H_3+H_4+\cdots +H_j +\cdots,
\end{equation}
where $H_j$ is a homogeneous polynomial of order $j$ with respect to  variables $\vx$. In our case the second term $H_2$ is as follows
\begin{equation}
\label{eq:Hlin}
H_2=\frac{p_1^2}{2\alpha\beta^2}+\frac{p_2^2}{2}+\frac{1}{2}(\beta^2+\delta)q_1^2-\beta q_1q_2+\frac{q_2^2}{2}.
\end{equation}
Since $H_2$ is quadratic form of $\vx$ it can be written as 
\begin{equation}
H_2=\frac{1}{2}\vx^T\hat \vH\vx,
\end{equation}
where $\hat \vH$ is  the  symmetric matrix
\begin{equation}
\hat \vH=\begin{pmatrix}
 \beta ^2+\delta  & -\beta  & 0 & 0 \\
 -\beta  & 1 & 0 & 0 \\
 0 & 0 & (\alpha  \beta ^2)^{-1} & 0 \\
 0 & 0 & 0 & 1 \\
\end{pmatrix}.
\end{equation}
The Hamilton equations  generated by $H_2$ are a linear system with constant coefficients of the following form
\begin{equation}
\label{eq:lin1}
\dot \vx=\vA\vx,\quad \text{where}\quad \vA=\vJ\hat\vH.
\end{equation}
Here $\vJ$ is the standard symplectic form satisfying $\vJ=-\vJ^T$.
 Considering  the eigenvalues of the matrix $\vA$  we can obtain information
 about the  stability of the linear system~\eqref{eq:lin1} near the equilibrium
 $\vx=0$. Following e.g.,~\cite{Markeev::78}, the necessary and sufficient
 condition for stability of linear Hamiltonian system is that the matrix $\vA$
 has  distinct and  purely imaginary eigenvalues, i.e., $\lambda_i=\rmi
 \omega_i, \ \lambda_{j+i}=-\rmi \omega_i, \ i=1,\dots j,$ and $\omega_i\in
 \field{R}$. 
In our case the characteristic polynomial of $\vA$ takes the form
\begin{equation}
\label{eq:char1}
p(\lambda)=\det[\vA-\lambda \vI]=\lambda^4+\frac{(\alpha+1)\beta^2+\delta}{\alpha\beta^2}\lambda^2+\frac{\delta}{\alpha\beta^2}.
\end{equation}
Because $p(\lambda)$ is an even function of $\lambda$ we can substitute $\sigma=\omega^2=-\lambda^2$, that gives
\begin{equation}
\label{eq:char11}
p(\sigma)=\sigma^2-\frac{(\alpha+1)\beta^2+\delta}{\alpha\beta^2}\sigma+\frac{\delta}{\alpha\beta^2}.
\end{equation}
It is easy to verify that the roots $\sigma_{1,2}$ of equation
$\eqref{eq:char11}$ are distinct real and positive only for $\delta>0$. This
implies that the  equilibrium  is linearly stable for $d>b$, and in this
paper we  restrict the value of $\delta$ to greater than zero.

\begin{figure}[t]
\begin{center}
\includegraphics[scale=0.7]{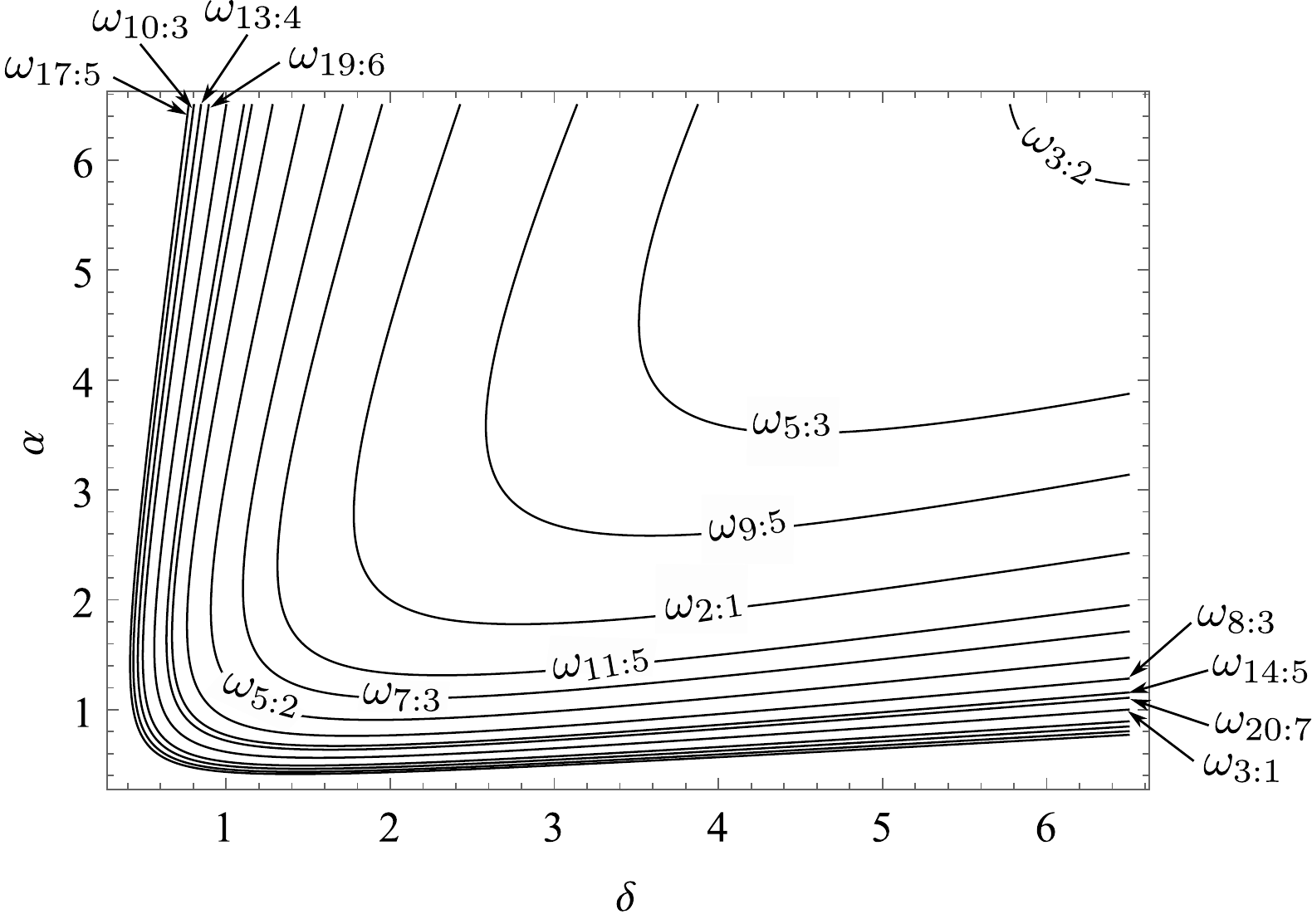}\caption{Examples of resonance curves plotted on the parameter plane for fixed $\beta=1$\label{fig:rez}}
\end{center}
\end{figure}

Next, we want to make the canonical transformation
\begin{equation}
\label{eq:my}
\vx=\vM\vX,\qquad \vM^T\vJ\vM=\vJ,
\end{equation}
such that the Hamiltonian $H_2$  in the new variables $\vX=(Q_1,Q_2,P_1,P_2)^T$ takes the form of the sum of two Hamiltonians for two independent harmonic oscillators
\begin{equation}
\begin{split}
\label{eq:HH2}
\cH_2=\frac{1}{2}\omega_1\left(Q_1^2+P_1^2\right)+\frac{1}{2}\omega_2\left(Q_2^2+P_2^2\right)=\frac{1}{2}\vX^T\hat \vK \vX,\quad \hat \vK=\diag(\omega_1,\omega_2,\omega_1,\omega_2).
\end{split}
\end{equation}
From equations ~\eqref{eq:lin1} and~\eqref{eq:my} we have the following condition for the matrix $\vM$
\begin{equation}
\vM\vJ\hat \vK=\vJ\hat \vH\vM.
\end{equation}
Then,  we look for  $\vM$ as a product $\vM=\vN\vL$, where $\vN$ transforms $\vA=\vJ\hat \vH$ into its Jordan form, and $\vL$ is  given by
\begin{equation}
\vL=\begin{pmatrix}
\rmi \field{I}_{2\times 2} & \field{I}_{2\times 2} \\
-\rmi \field{I}_{2\times 2}  & \field{I}_{2\times 2} \\
\end{pmatrix},
\end{equation}
where $\field{I}$ is the unit matrix.
Matrix $\vN$ is built from the eigenvectors of the matrix $\vA$ corresponding to eigenvalues  $\omega_{1,2}$. 
Choosing  appropriate order of  eigenvectors in $\vA$ as well as its
lengths we can obtain the real transformation $\vM=\vN\vL$ satisfying  the canonical
condition~$\vM^T\vJ\vM=\vJ$, see e.g.,~\cite{Markeev::78}. After such transformation $H_2$ takes the form~\eqref{eq:HH2}
with characteristic frequencies
\begin{equation}
\begin{split}
\label{eq:characteristic}
\omega_1&=\frac{\sqrt{2}\sqrt{(\alpha +1) \beta ^2+\sqrt{(\alpha
   +1)^2 \beta ^4-2 (\alpha -1) \beta ^2 \delta
   +\delta ^2}+\delta }}{2   \beta \sqrt{\alpha }
 }, \\
   \omega_2 &= \frac{\sqrt{2}\sqrt{(\alpha +1) \beta ^2-\sqrt{(\alpha
   +1)^2 \beta ^4-2 (\alpha -1) \beta ^2 \delta
   +\delta ^2}+\delta }}{2 \beta\sqrt{\alpha }
    }.
   \end{split}
\end{equation}
As we can notice these frequencies are different in general. However, it is easy
to verify that for some specific values of parameters $(\alpha,\beta,\delta)$
thy  become linearly dependent over the rational numbers. We say
that the eigenfrequences $\omega_{1,2}$ satisfy a resonance relation of order
$k$ if there exist integers $(m,n)$ such that
\begin{equation}
\label{eq:rez}
m\omega_1+n\omega_2=0,\qquad  |m|+|n|=k.
\end{equation}
 Figure~\ref{fig:rez} presents  examples of the  resonance curves plotted on the parameter  plane $(\delta,\alpha)$ for the fixed $\beta=1$. In this figure we use the notation
  \begin{equation}
  \label{omega}
  \omega_{n:m}:=\left\{(\alpha,\beta,\delta)\in \R\ \Big{|}\ \frac{\omega_1}{\omega_2} = \frac{n}{m}\right\}.
  \end{equation}
  In fact  we can obtain the explicit formulae for the resonance. Namely,
  substituting~\eqref{eq:characteristic} into~\eqref{eq:rez} and solving with
  respect to $\delta$, we obtain
 \begin{equation}
 \delta=\frac{\beta ^2 \left(\alpha +\alpha 
  l^4-2 l^2\pm\sqrt{\alpha }
   \left(l^2+1\right) \sqrt{\alpha
   +\alpha  l^4-2 (\alpha +2)
   l^2}\right)}{2 l^2},\quad l=\frac{n}{m}.
   \label{delta}
\end{equation}  
Limiting the  normalization of the Hamiltonian $H$~\eqref{eq:hamres1} only to the
quadratic part  $\scH_2$, does not give in general sufficient accuracy of  solutions of Hamilton's equations of the original untruncated $H$.  
Thus, in order to improve the accuracy we need to take into account the higher order terms
of $H$, and normalize them so that the Hamiltonian and the dynamics become
especially simple.  We can do this  by means of  a sequence of nonlinear
canonical transformations with some  appropriately chosen generating function
\begin{equation}
\label{eq:Sgen}
S=\vP^T\vq+W(\vq,\vP),\qquad W=W_3+\dots +W_K,
\end{equation}
where 
\[
W_K=\sum_{\nu_1+\dots +\mu_n=K}w_{\nu_1,\dots \mu_n}q_1^{\nu_1}\cdots q_n^{\nu_n}P_1^{\mu_1}\cdots P_n^{\mu_n},
\]
 for details consult e.g.,~\cite{Markeev::78,Jorba::99}. Let 
 \begin{equation}
 	\label{eq:23}
 \vQ=\pder{S}{\vP}=\vq+\pder{W(\vq,\vP)}{\vP},\qquad \vp=\pder{S}{\vq}=\vP+\pder{W(\vq,\vP)}{\vq}
 \end{equation}
 be the canonical transformation generated by~\eqref{eq:Sgen}. Then, from the implicit function theorem, in the neighborhood on the equilibrium $\vX=0$ we can express $(\vq,\vp)$ as a functions of $(\vQ,\vP)$. Namely, one can solve~\eqref{eq:23} for $\vq$ and $\vp$,
 treating the derivatives as known, and then recursively substitute
 those expressions into themselves. Because $W$ contains polynomials of
 degree 3 or higher, the Taylor series is recovered, with the first
 terms
\begin{equation}
 \vq =\vQ-\pder{W(\vQ,\vP)}{\vP}+\dots ,\qquad \vp=\vP+\pder{W(\vQ,\vP)}{\vQ}+\dots .
 \end{equation}
The Hamiltonian function  in the new variables is reduced to the
  Birkhoff normal form $\cH_K(\vX)$ of order $K$, i.e., with all homogeneous
  terms of degrees up to $K$ normalized so that they Poisson commute with the
  quadratic part
\begin{equation}
H(\vx)=\cH_K(\vX)+\cO(\vX^{K+1}),\quad \{\cH_K,\cH_2\}=0.
\label{eq:normals}
\end{equation}
Introducing the action-angle variables as the symplectic polar coordinates $(I_i,\phi_i)$ defined by
\[
 Q_i=\sqrt{2I_i}\sin\phi_i,\qquad P_i=\sqrt{2I_i}\cos\phi_i,\qquad i=1,\dots, n
\]
and discarding the non-normalized terms in~\eqref{eq:normals} we obtain the
integrable system whose Hamiltonian $\cH_K(\vI)$ depends only on actions and
whose trajectories will round the tori $\vI=\const$ with frequencies
$\vOmega=\partial_{\vI}\cH_K$. Since $\cH_K(\vX)$ consist of homogeneous terms of degrees up to  $K$  its Birkhoff normal form can be written in action-angle variables as a polynomial of degree $[K/2]$ in $\vI$. We can transform $H$ to such form by a sequence of canonical transformations provided eigenfrequences $\omega_i$ do not satisfy any resonance relation of order $K$ or less.

The Birkhoff normal form of degree four  is given by
\begin{equation}
\label{eq:fourth}
\cH_4=\omega_1 I_1+\omega_2 I_2+h_{20}I_1^2+h_{11}I_1I_2+h_{02}I_2^2,
\end{equation}
where the coefficients $h_{20},h_{11},h_{02}$ related to our system have the form
\begin{equation}
\begin{split}
h_{20}&=\frac{\omega _1^2 \omega _2^4 \left(\omega _1^2-1\right)
   \left(3 \delta  \left(\omega
   _1^2-1\right) \left(\omega
   _2^2-1\right)+\left(3 \omega _2^2+1\right)
   \omega _1^2-3 \omega _2^2+3\right)}{16 \delta
   ^2 \left(\omega _1^2-\omega _2^2\right)^2
   \left(\omega _2^2-1\right)},\\
   h_{11}&=\frac{\omega _1^3 \omega _2^3 \left(\left(1-3 \omega _2^2\right)
   \omega _1^2+\omega _2^2-3-3 \delta 
   \left(\omega _1^2-1\right) \left(\omega
   _2^2-1\right)\right)}{4 \delta ^2
   \left(\omega _1^2-\omega _2^2\right)^2},\\
   h_{02}&=\frac{\omega _1^4 \omega _2^2 \left(\omega
   _2^2-1\right) \left(3 \delta  \left(\omega
   _1^2-1\right) \left(\omega _2^2-1\right)+3
   \left(\omega _2^2-1\right) \omega _1^2+\omega
   _2^2+3\right)}{16 \delta ^2 \left(\omega
   _1^2-1\right) \left(\omega _1^2-\omega
   _2^2\right)^2}.
\end{split}
\end{equation}
 \begin{figure}[t]
\centering
\subfigure[Numerical]{  
\includegraphics[width=0.48\textwidth]{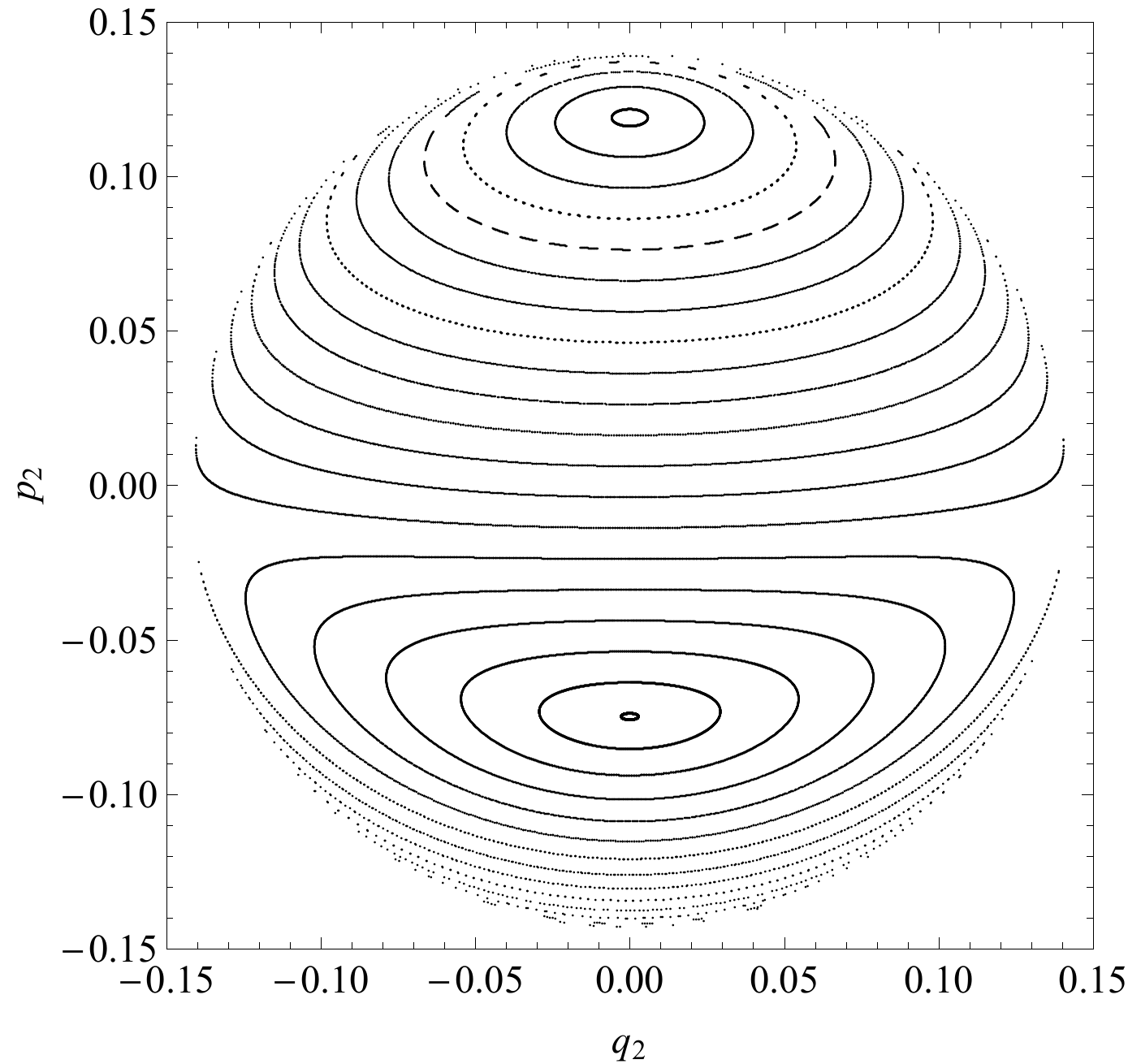} 
    }
    \subfigure [Analytical]{
\includegraphics[width=0.48\textwidth]{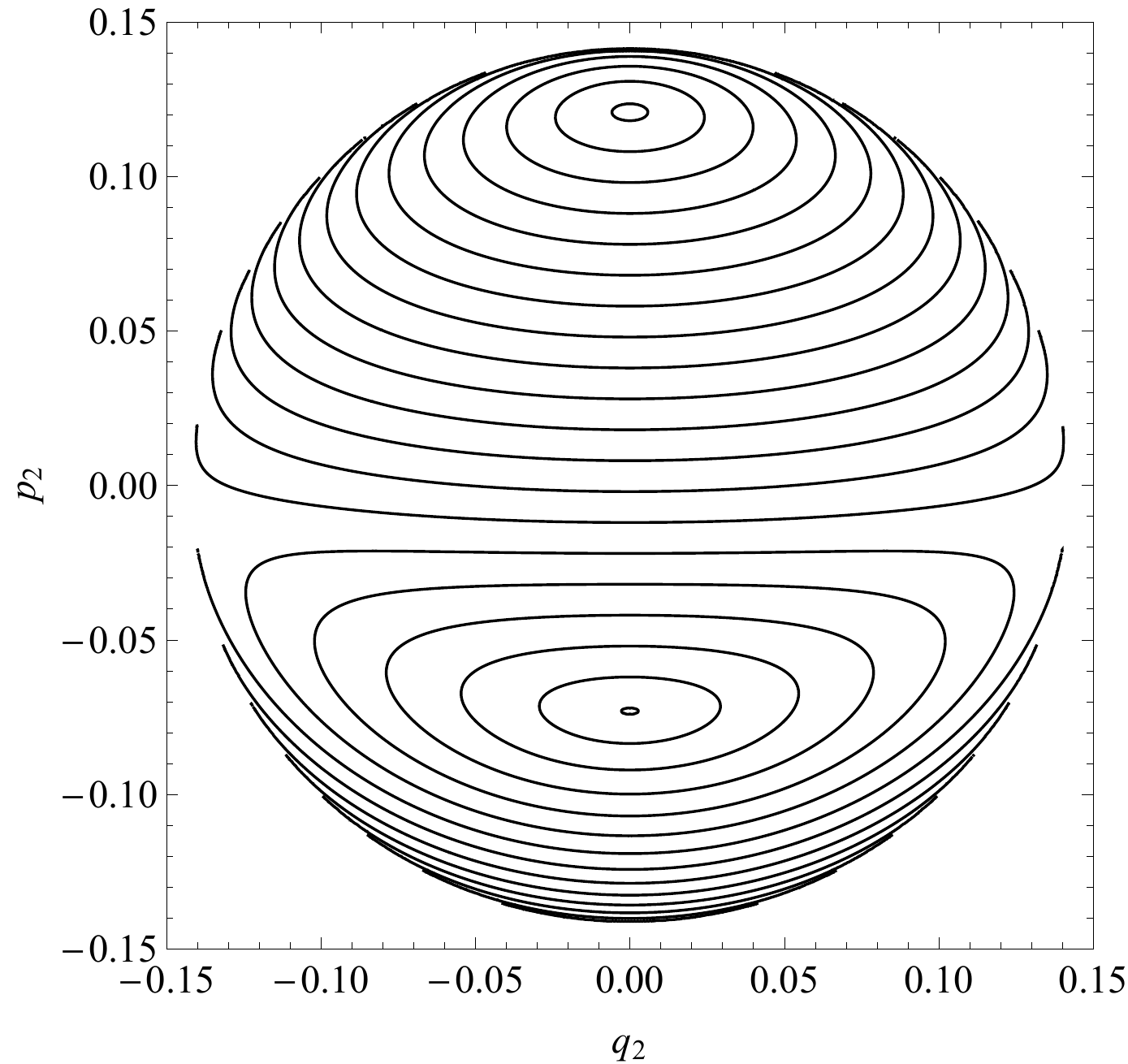}
}
\caption{Poincar\'e sections on the surface $q_1=0$, with $p_1>0$\label{fig:poin_33}}
\end{figure}
\begin{figure}[h!]
\begin{center}
\includegraphics[scale=0.6]{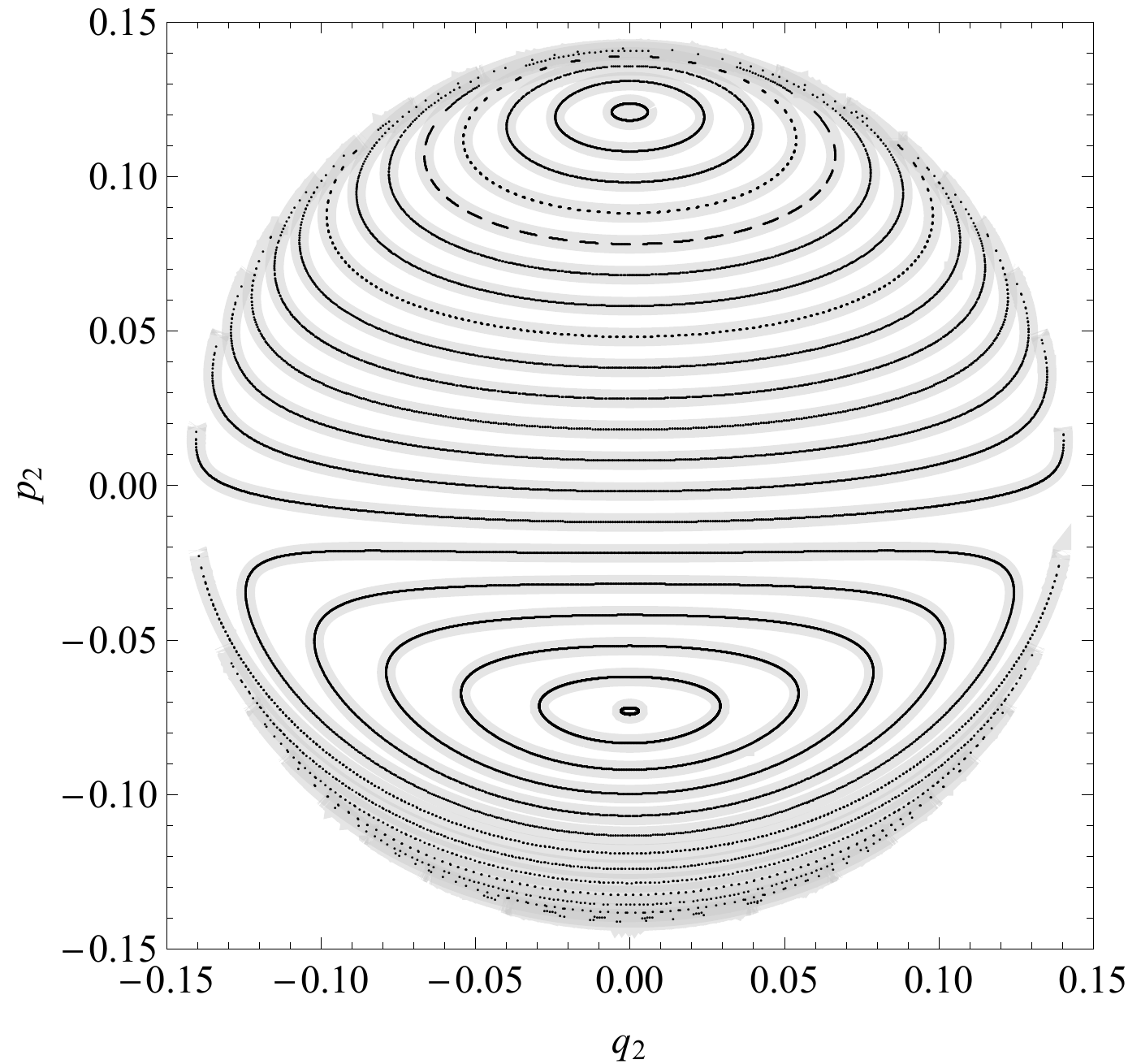}
\caption{Superposition of Figs.~\ref{fig:poin_33}(a)-(b)\label{fig:poin_4} }
\end{center}
\end{figure}
Now let us try to deduce some interesting information from the normalized
Hamiltonian~\eqref{eq:fourth}.
First  of all, in order to check that the normalization up to order four gives
sufficiently good  accuracy,  we have to  compare solutions of Hamilton equations governed by Hamiltonian~\eqref{eq:fourth} with~\eqref{eq:hamres1}. We can do this easily for example by comparison of 
Poincar\'e cross sections for original Hamiltonian system and its normal form of degree four. Let us note that action variables $(I_1,I_2)$ are the first integrals of the normalized Hamiltonian $\cH_4$.  We can chose one of
them, for example $I_1$ and make the inverse canonical transformation in
order to go back to the original variables $(q_1,q_2,p_1,p_2)^T$. Then for the chosen
energy level we make the contour plot of $I_1$ restricted to the
plane $(q_2,p_2)$ with  $q_1=0$. Figure~\ref{fig:poin_33} presents numerical and analytical
Poincar\'e cross sections constructed  for  chosen values of parameters belonging to the stability region, namely:
$
\omega_1=\omega_1=1.75, \ \omega_2=0.5, \ \delta=0.75,
$
 with cross-section plane $q_1=0$ and $p_1>0$, on the energy level
 $E=E_{\text{min}}+0.01$. As expected, for $E$ close to the energy minimum corresponding to an equilibrium 
 both the images are very regular. In fact each of them can be divided into two
 regions filled by invariant tori around two stable particular periodic
 solutions.  As we can notice, the differences between numerical and analytical
 computations are not visible. See especially the Figure~\ref{fig:poin_4}
 showing the superposition of Figures~\ref{fig:poin_33}(a)-(b), where for
 better readability the analytical loops have been plotted  in bold gray lines.
 
\begin{figure}[t]
\begin{center}
\includegraphics[scale=0.65]{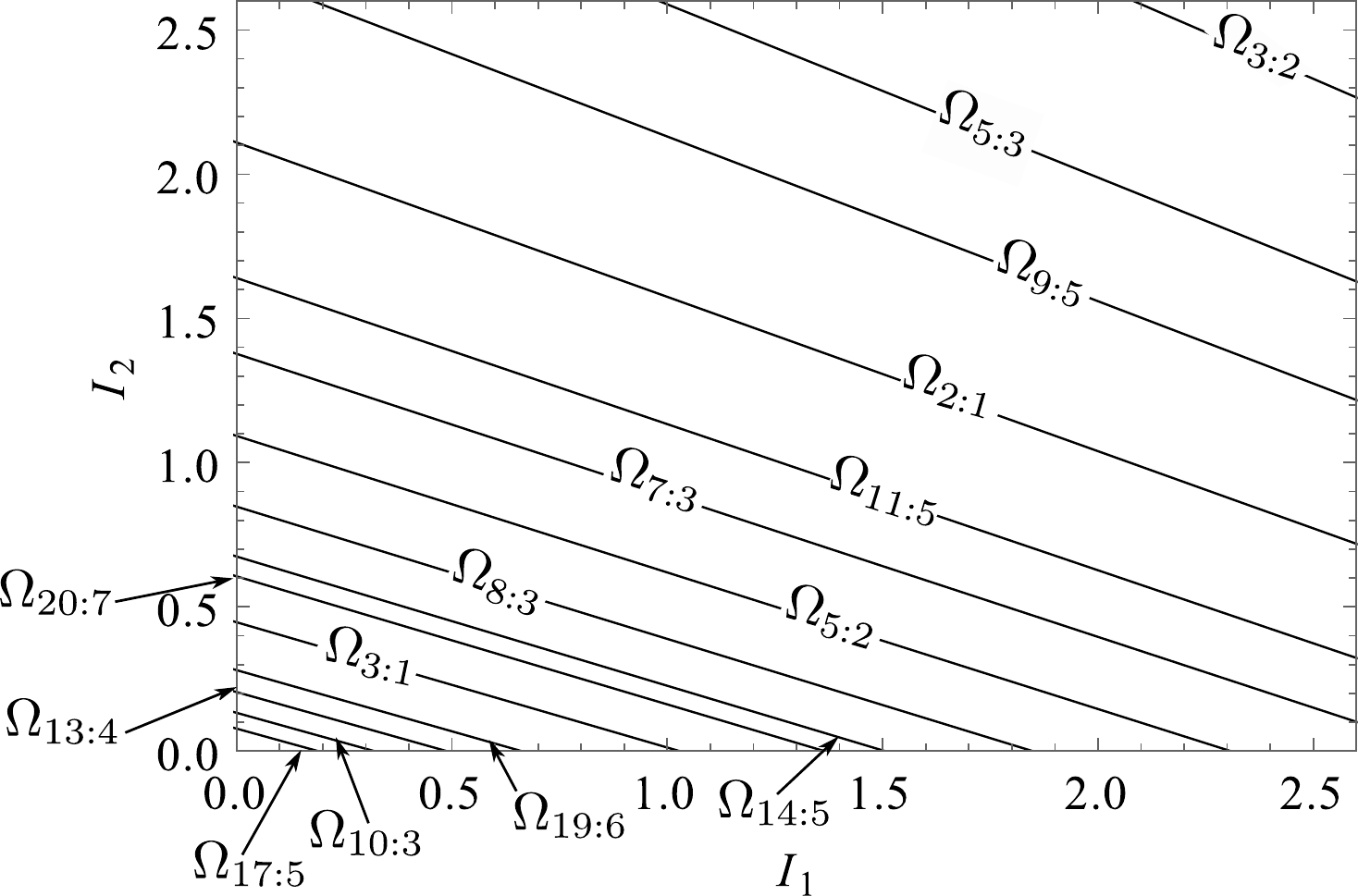}
\caption{
	Contour plot showing  the resonance curves  on the $(I_1, I_2)$ plane for the fixed values of  parameters: $\omega_1=1.75, \ \omega_2=0.5, \ \delta=0.75$\label{fig:rez11}}
\end{center}
\end{figure}

\begin{figure}[h!]
	\begin{center}
		\includegraphics[scale=0.7]{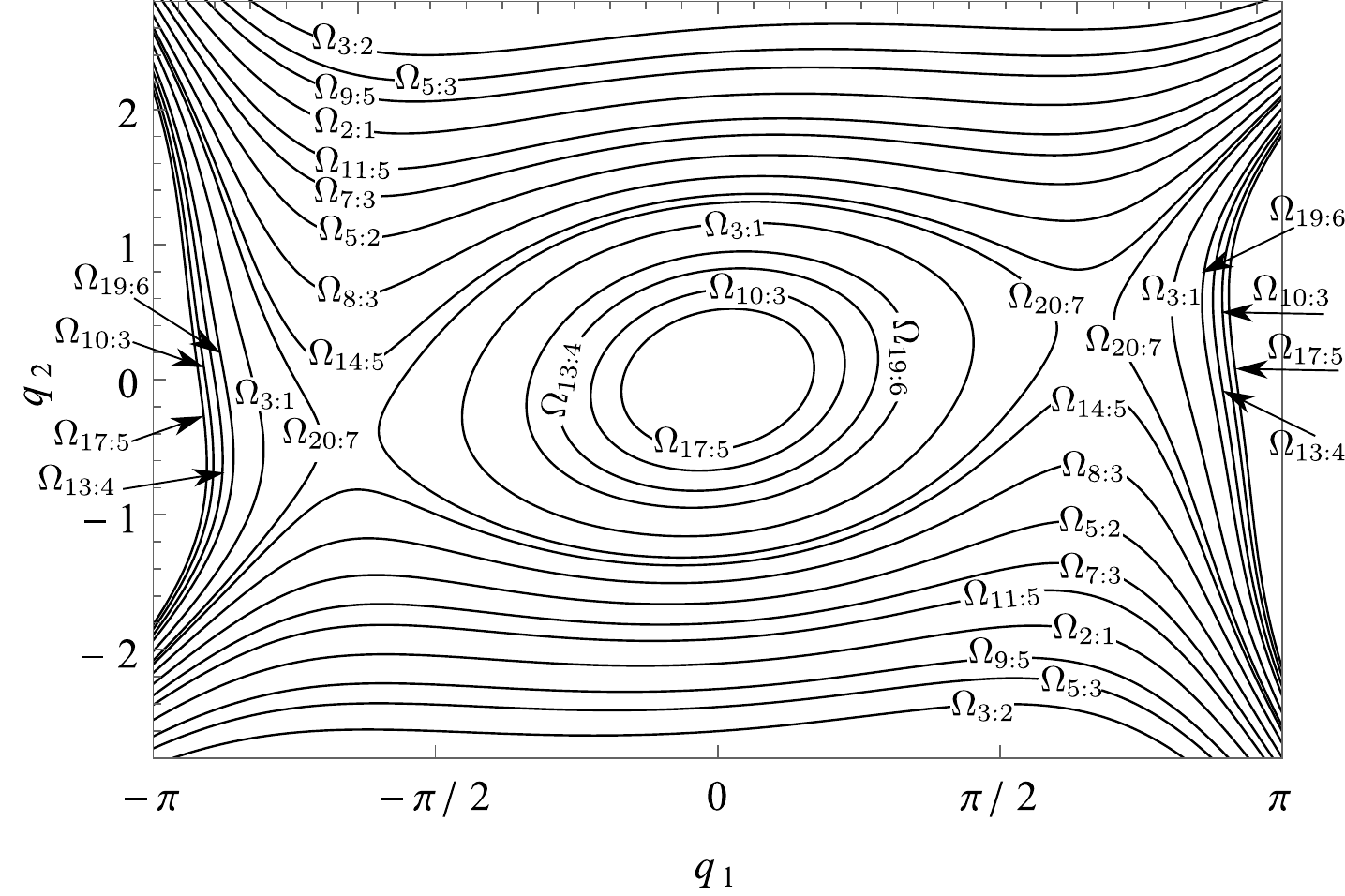}
		\caption{	Contour plot showing  the resonance curves  on the $(q_1, q_2)$ plane with $p_1=p_2=0$ for the fixed values of  parameters: $\omega_1=1.75, \ \omega_2=0.5, \ \delta=0.75$\label{fig:rez2}}
	\end{center}
\end{figure}

\begin{figure}[h!]
	\vskip -10pt
	\begin{center}
		\includegraphics[scale=0.7]{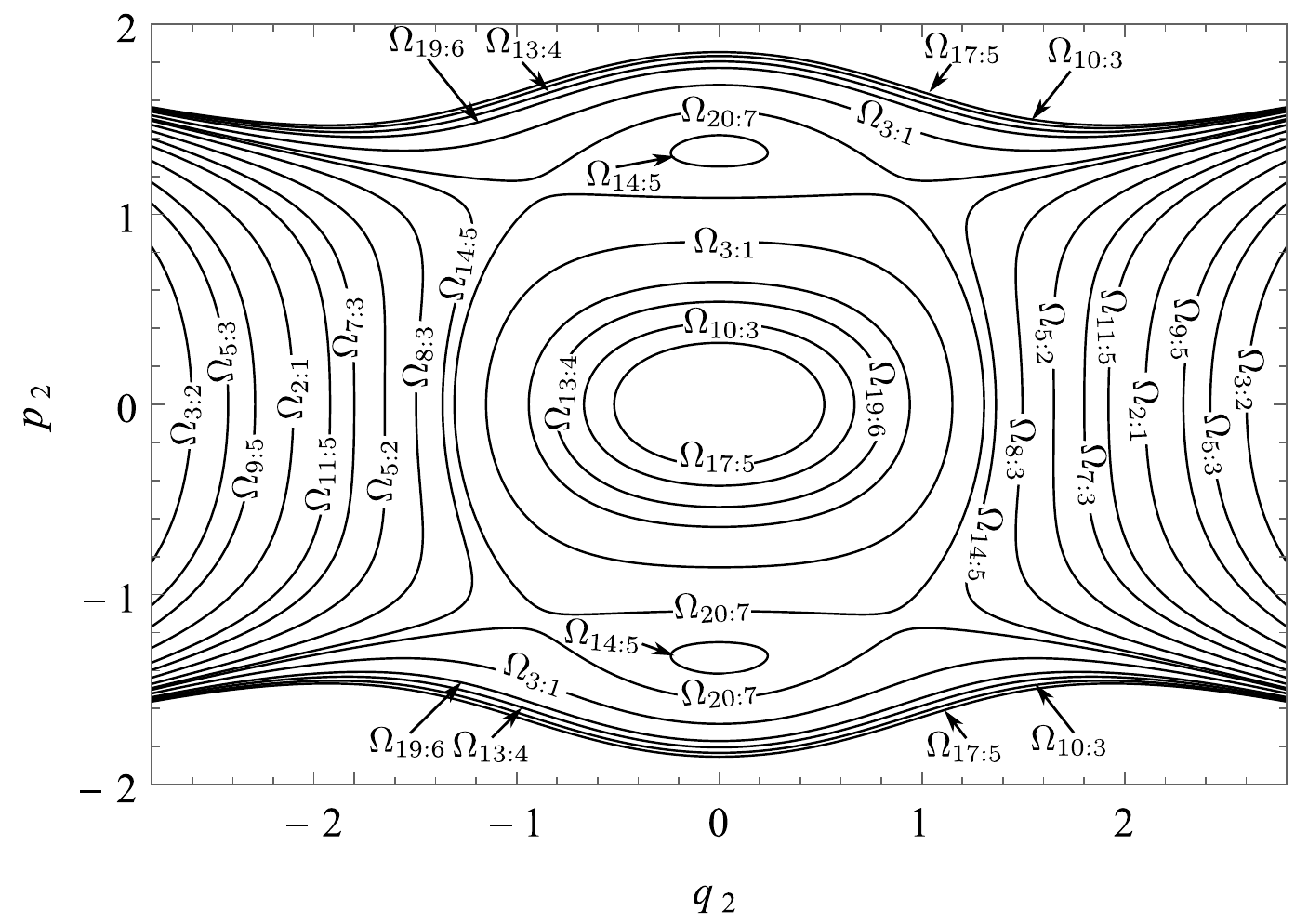}
		\caption{	Contour plot showing  the resonance curves  on the $(q_2, p_2)$ plane with $q_1=p_1=0$ for the fixed values of  parameters: $\omega_1=1.75, \ \omega_2=0.5, \ \delta=0.75$ \label{fig:rez3}}
	\end{center}
\end{figure}

\begin{figure}[h]
\begin{center}
\includegraphics[scale=0.52]{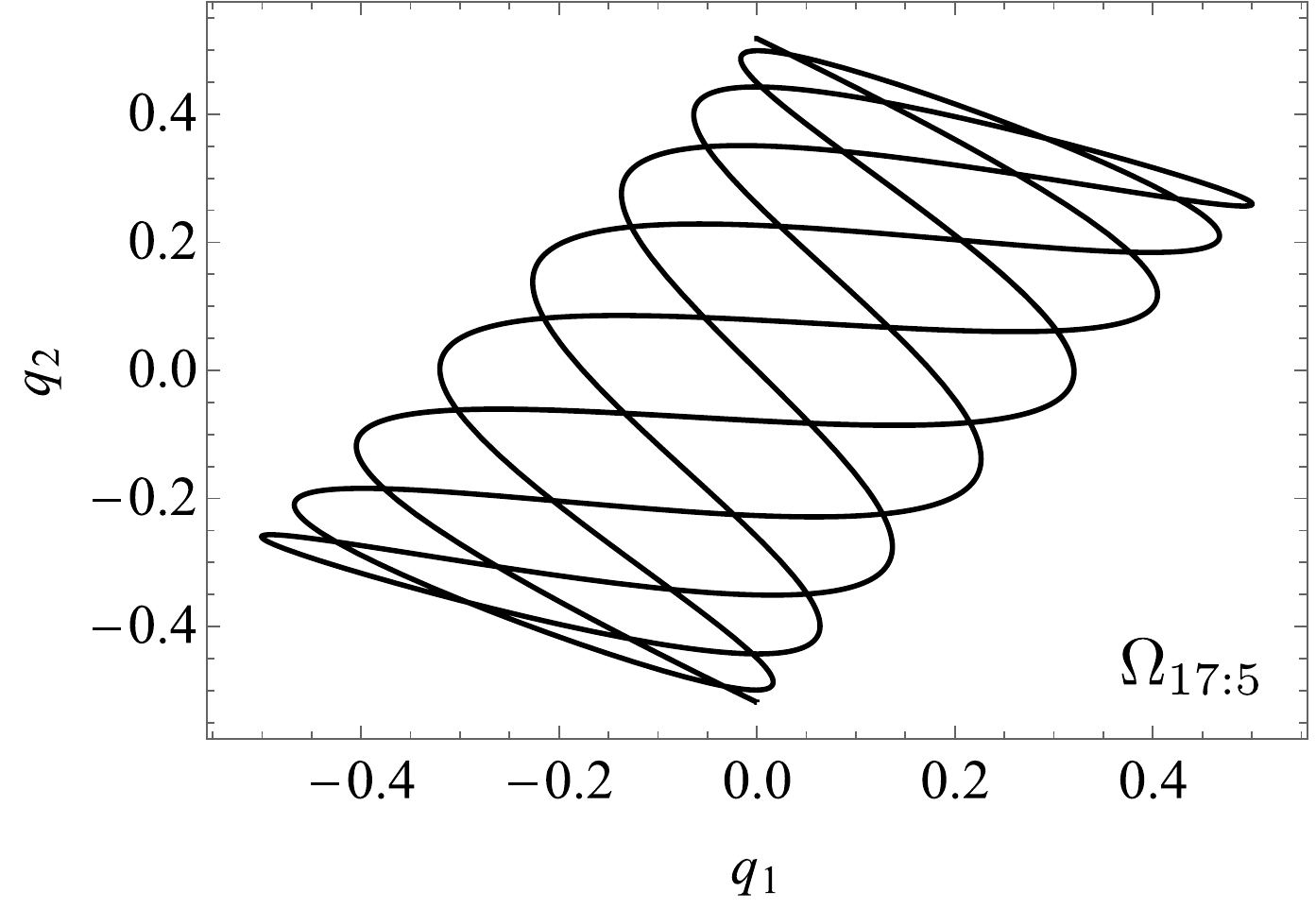}
\includegraphics[scale=0.52]{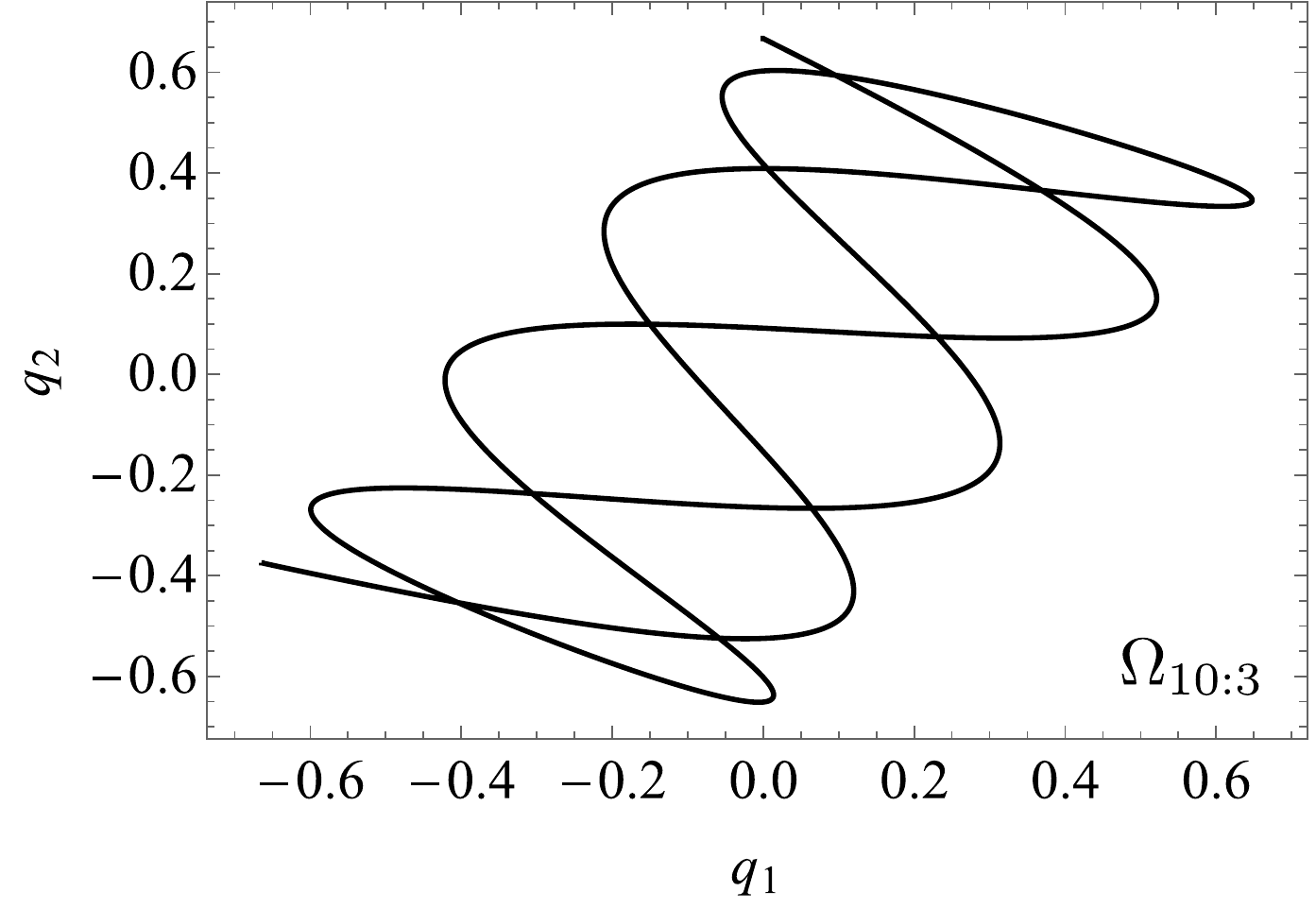}
\caption{Examples of trajectories in the configuration space given by  the numerical computations with the initial values related to certain resonances presented in Figs.~\ref{fig:rez2}-\ref{fig:rez3}\label{fig:lisa}}
\end{center}
\end{figure}

 As we mentioned previously, the Birkhoff normalization can be also very effective in finding families of  periodic solutions. 
 Figures~\ref{fig:rez11}-\ref{fig:rez3} present contour plots showing examples
 of such families
 \begin{itemize}
  \item on the  $(I_1, I_2)$ plane,
  \item on the $(q_1, q_2)$ plane with $p_1=p_2=0$,
  \item on the $(q_2,p_2)$ plane with $q_1=p_1=0$,
 \end{itemize}
respectively, where $(\Omega_1,\Omega_2)$  are defined by
\begin{equation}
\begin{split}
\Omega_1=\pder{\cH_4}{I_1}&= \omega_1+2h_{20}I_1+h_{11}I_2,\quad
\Omega_2=\pder{\cH_4}{I_2}=\omega_1+2h_{02}I_2+h_{11}I_1,\\ &
\quad \text{and}\quad \Omega_{n:m}:=\left\{(I_1,I_2)\in \R^2 \ \Big{|}\ \frac{\Omega_1}{\Omega_2}=\frac{n}{m} \right\}.\end{split}
\end{equation}
These figures are very helpful because from them we can read initial conditions
conditions for which the motion of the system is periodic. For example,
Figure~\ref{fig:lisa} presents  periodic orbits in the configuration space given by  the numerical computations with the initial values related to resonances $\Omega_{17:5}$ and $\Omega_{10:3}$ respectively. 
\section{Conclusions}
As we have seen, in order to determine stability of equilibrium solution,
detect families of periodic solutions as well as find approximation of analytic
solutions of equations of motion  the Birkhoff normalization proves very useful.
However, it is worth mentioning certain inconveniences associated with
this method.
Namely,   it  gives us opportunity to investigate behavior of the system over large time intervals, with sufficiently good accuracy, only in the neighborhood
of equilibrium. Furthermore, it should be emphasized that the linear stability of the equilibrium $\vx=0$ does not imply its stability in the  Lyapunov sense. This is due to the fact that  the
discarded parts of the series can destroy the stability in the long timescale.
It would seem that the higher degree normalization should gives us a
better approximation of reality. However, in general there does not exist a convergent  Birkhoff
transformation, see e.g.,~\cite{Perez::03}, so the estimation of those terms is not straightforward. To check the non-linear stability of equilibrium for Hamiltonian vector field $\vv_H(\vx)$ more involved analysis is necessary, e.g., application
of the second Lyapunov method or the Arnold-Moser theorem, which
itself relies on normal form, see~\cite{Merkin::97}.

Despite these limitations, the Birkhoff normalization is still a very useful source
of important information about dynamics and often the starting point of further analysis.
\section*{Acknowledgement}
The work has been supported by grants No. DEC-2013/09/B/ST1/ 04130 and DEC-2011/02/A/ST1/00208  of National Science Centre of Poland.
%
%
%

\end{document}